\documentclass[referee,sn-mathphys]{sn-jnl}

\jyear{2021}%

\begin{document}

\title[A Publicly Available Dataset of Out-of-Field Dose Profiles]{A Publicly Available Dataset of Out-of-Field Dose Profiles of a 6 MV Linear Accelerator}


\author*[1,2]{\fnm{Samuel C.} \sur{Peet}}\email{samuel.peet.physics@gmail.com}
\author[1]{\fnm{Naasiha} \sur{Cassim}}
\author[1,2,3]{\fnm{Tanya} \sur{Kairn}}
\author[2]{\fnm{Jamie V.} \sur{Trapp}}
\author[1,2,3]{\fnm{Scott B.} \sur{Crowe}}

\affil[1]{\orgname{Royal Brisbane and Women's Hospital}, \orgaddress{\city{Herston}, \postcode{4029}, \state{QLD}, \country{Australia}}}
\affil[2]{\orgname{Queensland University of Technology}, \orgaddress{\city{Brisbane}, \postcode{4001}, \state{QLD}, \country{Australia}}}
\affil[3]{\orgname{University of Queensland}, \orgaddress{\city{Brisbane}, \postcode{4072}, \state{QLD}, \country{Australia}}}

\abstract{An increase in radiotherapy-induced secondary malignancies has led to recent developments in analytical modelling of out-of-field dose. These models must be validated against measurements, but currently available datasets are outdated or limited in scope. This study aimed to address these shortcomings by producing a large dataset of out-of-field dose profiles measured with modern equipment. A novel method was developed with the intention of allowing physicists in all clinics to perform these measurements themselves using commonly available dosimetry equipment. A standard 3D scanning water tank was used to collect 36 extended profiles. Each profile was measured in two sections, with the inner section measured with the beam directly incident on the tank, and the outer section with the beam incident on a water-equivalent phantom abutted next to the tank. The two sections were then stitched using a novel feature-matching approach. The profiles were compared against linac commissioning data and manually inspected for discontinuities in the overlap region. The dataset is presented as a publicly accessible comma separated variable file containing off-axis ratios at a range of off-axis distances. This dataset may be applied to the development and validation of analytical models of out-of-field dose. Additionally, it may be used to inform dose estimates to radiosensitive implants and anatomy. Physicists are encouraged to perform these out-of-field measurements in their own clinics and share their results with the community.}


\maketitle

\section{Introduction}

Improvements in diagnostic and treatment technologies have resulted in overall cancer survival rates increasing in recent times \cite{miller2019}. However, this increase in radiation therapy survivorship has coincided with a commensurate increase in the rate of radiation induced secondary malignancies later in life \cite{dracham2018}. In response, there has been increasing effort to better understand out-of-field doses delivered during radiation therapy treatments \cite{kry2017}. Beyond secondary cancer induction, accurate out-of-field dose estimates are also necessary for assessing doses to radiosensitive implants such as cardiac devices \cite{miften2019, peet2016} and radiosensitive anatomy as in the case of pregnant patients \cite{stovall1995, peet2021}. There is also a need for increasingly accurate out-of-field dose estimates to inform epidemiological studies \cite{harrison2017}.

Methods of estimating out-of-field doses include consulting simple reference data in the literature \cite{stovall1995,mutic1999,ruben2011,miljanic2014}, Monte Carlo simulations \cite{kry2006,kry2007,bednarz2009}, and increasingly refined analytical models \cite{benadjaoud2012,taddei2013,schneider2019,wilson2020}. These models may need to consider the physical geometry of clinical linacs, including jaws, primary collimator, MLC leaves and carriage, and the arrangement of additional head shielding. This is further complicated by the relative positions of the linac treatment head and patient, such as during non-coplanar cranial radiotherapy \cite{kairn2019}. Modelling and verifying the effects of these factors would require the measurement of a large number of out-of-field profiles with many collimator orientations. As out-of-field dose modelling matures, it is also reasonable to expect that computational models will augment traditional dose calculation algorithms in commercial treatment planning systems. Commissioning these models may require the measurement of profiles much farther outside the field than physicists are currently accustomed. 

Available datasets in the literature generally contain older model linacs and MLCs that are rarely seen today \cite{stovall1995,mutic1999} and present coarse resolution dose profiles in a limited set of collimator orientations. Physicists wishing to use these profiles are forced to interpolate values from the printed figures by hand. This is a sub-optimal approach, and a contemporary solution using modern dosimetry equipment should achieve much greater spatial resolution while also presenting the data digitally in a manner that is computationally digestible.

In this article we establish a method of measuring high-quality out-of-field dose profiles using typically available clinical physics equipment. This method is then followed to produce a comprehensive and publicly available dataset of out-of-field dose profiles of our clinical linac.

\section{Methods}

\subsection{Equipment}

The linac under investigation in this study was a Varian Clinac iX (Varian Medical Systems, Palo Alto, USA) with a Millennium 120 MLC. All measurements were performed using the 6 MV photon energy with 600 MU/min dose rate and flattening filter. A PTW Semiflex 0.3 cm$^3$ 31013 (PTW-Freiburg, Freiburg, Germany) ionisation chamber was used for all profile scans. The dimensions of this chamber afforded lower volume averaging compared to larger thimble chambers, while still maintaining the sensitivity necessary for far out-of-field measurements. The chamber was affixed to a PTW BEAMSCAN water tank using the TRUFIX chamber positioning system and steered with version 4.3 of the control software. The BEAMSCAN water tank had a usable scanning range of 500 mm (horiz.) $\times$ 500 mm (horiz.) $\times$ 415 mm (vert.), and 15 mm thick PMMA walls. A PTW Semiflex 3D 0.07 cm$^3$ 31021 ionisation chamber was used as a reference to correct for fluctuations in the beam output during each scan. The chamber voltages were set to 400 V as per manufacturer recommendations, and the integrated water tank electrometer was set to low range. The control software was programmed to measure profiles in continuous scanning mode with a chamber speed of 5 mm/s and data points reported every 2 mm. A 30 $\times$ 30 $\times$ 40 cm$^3$ stack of Virtual Water (Standard Imaging, Middleton, USA) was used as additional scattering material.

\subsection{Measurement Setup}

Each profile was measured as a piece-wise combination of two scans to construct a much longer profile measurement than a typical water tank would allow. The first section acquired the in-field and near out-of-field region of each profile, while the second section captured the far out-of-field region. An overlap area of approximately 15 cm was included in each pair of profile sections.

\begin{figure}[h]
\centering
\includegraphics[width=0.6\textwidth]{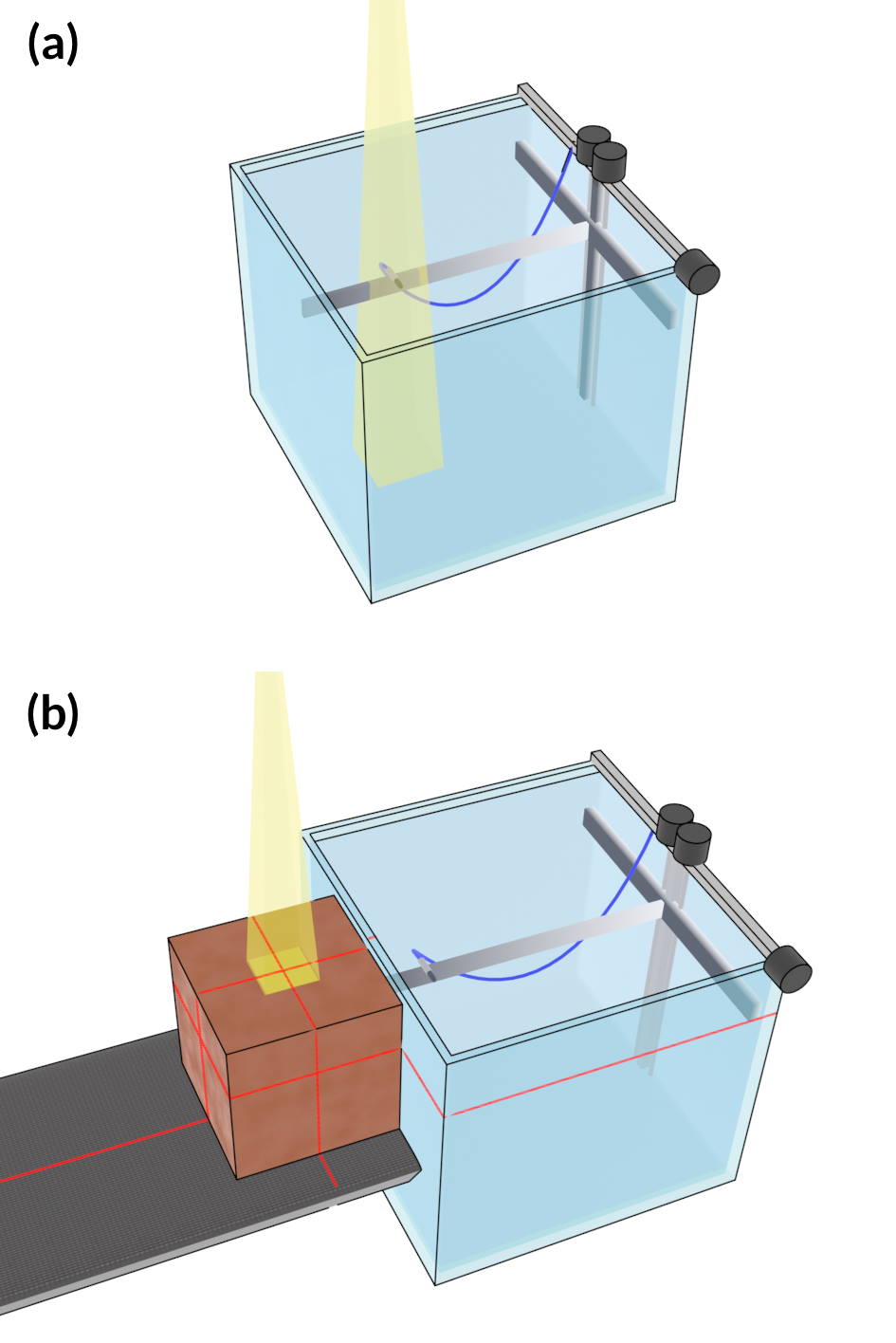}
\caption{Equipment setup for the measurement of profiles in this study. (a) Inner profile section: The field is directed into the tank near the wall. (b) Outer profile section: The tank is shifted outside of the field and abutted to a stack of Virtual Water to ensure full scattering conditions.}\label{fig:tank_setup} 
\end{figure}

The first measurement geometry can be seen in Figure \ref{fig:tank_setup}a. To begin, the tank was positioned such that the long axis of the chamber was orthogonal to the scanning direction, with a source to surface distance of 90 cm. The tank was then translated such that the central axis of the beam was 15 cm from the maximum chamber travel position on one side, and 35 cm from the maximum chamber travel position on the opposite side. This allowed the inner section of each profile to be collected out to 35 cm from the central axis while still allowing adequate scatter around the field.

The measurement geometry was then altered to capture the far out-of-field section of each profile, as seen in Figure \ref{fig:tank_setup}b. The tank was translated 31.5 cm in the direction of the profile, and a stack of Virtual Water was positioned in the field at 90 cm source to surface distance to re-establish full scattering conditions. Care was taken to ensure that the Virtual Water was abutted firmly against the tank wall with minimal air gaps. 

\subsection{Post-processing}

\begin{figure}[h]
\centering
\includegraphics[width=1.0\textwidth]{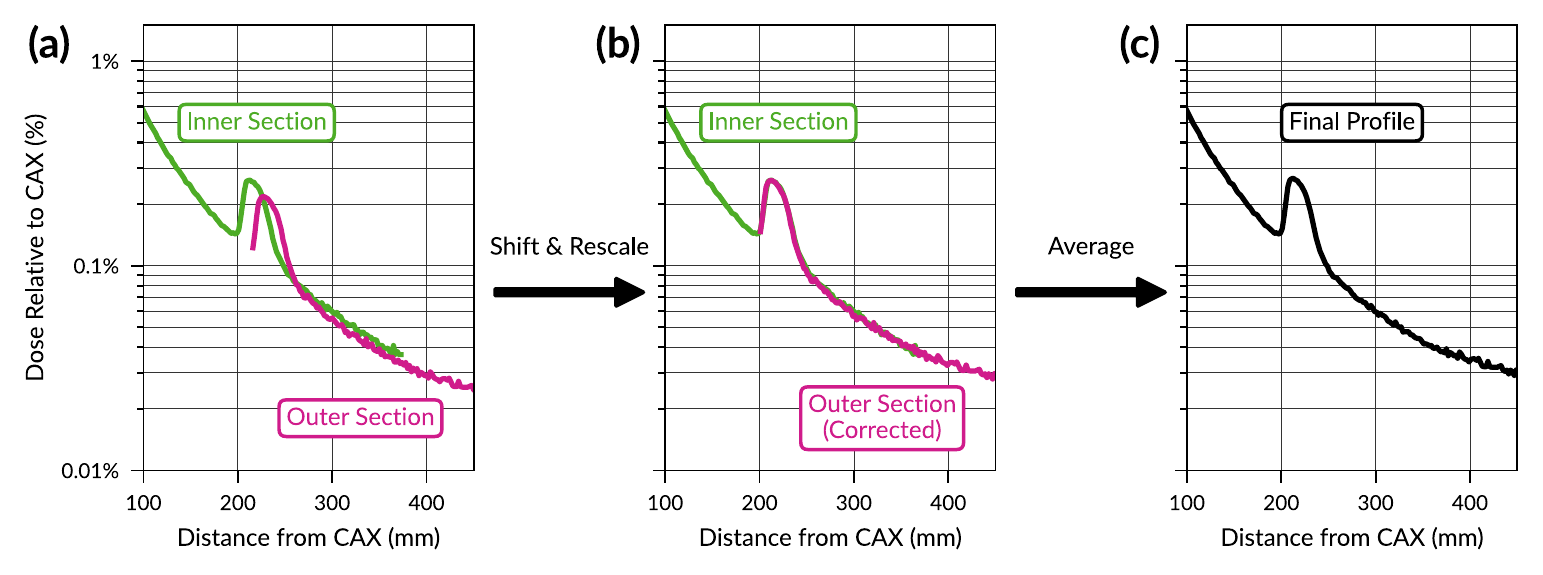}
\caption{The process of stitching two profile sections to make a whole. (a) The two raw profile sections with a strong feature for matching. (b) The outer section shifted such that the features overlap. (c) The two sections conjoined, with the average value taken in the overlap region.}\label{fig:feature_matching} 
\end{figure}

The two sections of each profile were stitched together to create a whole. This was non-trivial, as the overlap regions did not necessarily coincide due to small air gaps between the Virtual Water and the tank wall, the non-water-equivalence of the tank wall, and positioning errors in the tank. To overcome this issue, each overlap region was searched for a prominent feature common to both sections of the profile. The far out-of-field section was then progressively shifted in 1 mm increments and re-scaled until the identified feature matched the inner profile section as well as possible. The two sections were then combined into one full profile, with an average being taken in the overlap region. A visual example of this process can be seen in Figure  \ref{fig:feature_matching}. No additional smoothing or filtering was applied. The full profiles were validated by comparisons with commissioning data and visual inspections for any discontinuities.

\subsection{Summary of Collected Profiles}

In total, 36 profiles were collected (see summary in Table \ref{tab:tank_profiles}). Profiles were gathered at three nominal field sizes: 5 $\times$ 5 cm$^2$, 10 $\times$ 10 cm$^2$, and 15 $\times$ 15 cm$^2$. Profiles in both the X and Y directions were measured with the jaws defining the field and the MLC in a 'parked' state (leaf tips fully retracted to approximately 21 cm off axis). These were repeated with the MLC defining the field and the jaws at the Varian recommended positions (X jaws 8 mm retracted from nominal field edge, Y jaws 2 mm retracted from nominal field edge). All profiles were measured with a source to surface distance of 90 cm and depth of 10 cm, except for the 10 $\times$ 10 cm$^2$ MLC defined profiles which were additionally measured at depths of 15, 20, and 25 cm. Every profile was measured twice: once aligning with the radial (gun-target) direction and once aligning with the transverse (left-right) direction. The collimator angle was set appropriately to achieve this, for example, a transverse X profile required collimator 0$^{\circ}$, while a radial X profile required collimator 90$^{\circ}$ (or 270$^{\circ}$). The radial direction corresponds to the cranio-caudal axis of a patient receiving an coplanar treatment with no couch rotation and the transverse direction corresponds to the cranio-caudal axis of a patient receiving a treatment with a 90$^{\circ}$ couch rotation. The radial and transverse profiles were collected in two separate measurement sessions as they required a complete reorientation of the tank. In total, the measurement time was approximately eight hours, including equipment setup/cleanup and data collection.

\begin{table}[h]
\begin{center}
\caption{Field configurations measured in this study. For each row in the table, X and Y profiles were gathered in both the radial and transverse orientations}\label{tab:tank_profiles}
\vspace*{2ex}
\begin{tabular} {c l l c}
&&&\vspace{-2mm} \\
Field Size (cm$^2$) & MLC State \hspace{8mm} & Symmetric Jaw Apertures (cm)    & Depth (cm)    \\
&&&\vspace{-2mm} \\
\hline
&&& \\
5 $\times$ 5        & Retracted         & (X = 5, Y = 5)        & 10            \\
5 $\times$ 5        & Defining field    & (X = 6.6, Y = 5.4)    & 10            \\
&&& \\
\hline
&&& \\
10 $\times$ 10      & Retracted         & (X = 10, Y = 10)      & 10            \\
10 $\times$ 10      & Defining field    & (X = 11.6, Y = 10.4)  & 10            \\
10 $\times$ 10      & Defining field    & (X = 11.6, Y = 10.4)  & 15            \\
10 $\times$ 10      & Defining field    & (X = 11.6, Y = 10.4)  & 20            \\
10 $\times$ 10      & Defining field    & (X = 11.6, Y = 10.4)  & 25            \\
&&& \\
\hline
&&& \\
15 $\times$ 15      & Retracted         & (X = 15, Y = 15)      & 10            \\
15 $\times$ 15      & Defining field    & (X = 16.6, Y = 15.4)  & 10            \\
&&& \\
\hline
\end{tabular}
\end{center}
\end{table}

\section{Results}

\subsection{Dataset Accessibility and Usage Notes}

The dataset had been made publicly accessible through the Zenodo platform and released under a Creative Commons Attribution 4.0 licence \cite{peet2021dataset}. The dataset consists of a single comma separated variable (CSV) file containing 37 columns, with the first column storing the off-axis distance of each measurement point, and the remaining 36 columns storing the off-axis ratios at these points for each profile. The first row contains a label for each profile with the format `[$field\_size$] [$mlc\_state$] [$depth$] [$direction$] [$orientation$]' where
\begin{itemize}
    \item $field\_size$ is the nominal field size at the isocentre,
    \item $mlc\_state$ designates the position of the MLC, taking the value `MLC defined' when the MLC defines the field, and `MLC parked' when the MLC is fully retracted,
    \item $depth$ is the depth of measurement,
    \item $direction$ indicates the direction of the profile, either X or Y, and
    \item $orientation$ indicates whether the measurement orientation was radial (GT) or transverse (LR).
\end{itemize}
For example, a profile under the Y jaw of a 10 $\times$ 10 cm$^2$ MLC defined field measured at a depth of 10 cm in the radial direction would have the label '10 x 10 MLC defined Y d10cm (GT)'. Each profile was normalised to the value at the central axis at the time of measurement so no further processing is needed to recover off-axis ratios. Users should be aware that the profiles are of unequal length, depending on whether they were acquired in the radial or transverse measurement orientation. An exploration of the dataset will be undertaken in the Discussion section.

\subsection{Dataset Validation}

In order to establish that the irradiation conditions and measurement setup were representative of normal practice, the inner component of the 10 $\times$ 10 cm$^2$ X and Y profiles were compared against the same profiles gathered during commissioning of the linac. A gamma comparison of the in-field sections, bounded by the 50\% isodose lines, showed 100\% agreement with criteria of 1\% dose difference (local normalisation) and 1 mm distance-to-agreement. This gave good confidence that the measurement equipment was set up correctly, that the linac was behaving nominally, and that ultimately the data gathered in the session was representative of clinical practice.

During the post-processing step, the outer section of each profile was shifted to coincide with the inner section of the profile. Given that the equipment was not moved during the measurement of all outer profile sections within a given orientation (radial or transverse), it follows that the ideal shift should be the same across all profiles. This was indeed found to be the case with the largest difference in ideal shifts being 1 mm. After combining the profile halves and taking the average in the overlap region, all full profiles were then manually inspected to ensure that there were no obvious discontinuities in the overlap region. None were found.
\section{Discussion}

\subsection{An Exploration of Dose Outside the Treatment Field}

This investigation produced a large dataset with many curious features. In this section we compare and contrast a number of profiles that illuminate the underlying geometry and radiation interactions. However, these are just a representative selection of the profiles, and we encourage interested readers to further examine the dataset for themselves.

\begin{figure}[h]
\centering
\includegraphics[width=1.0\textwidth]{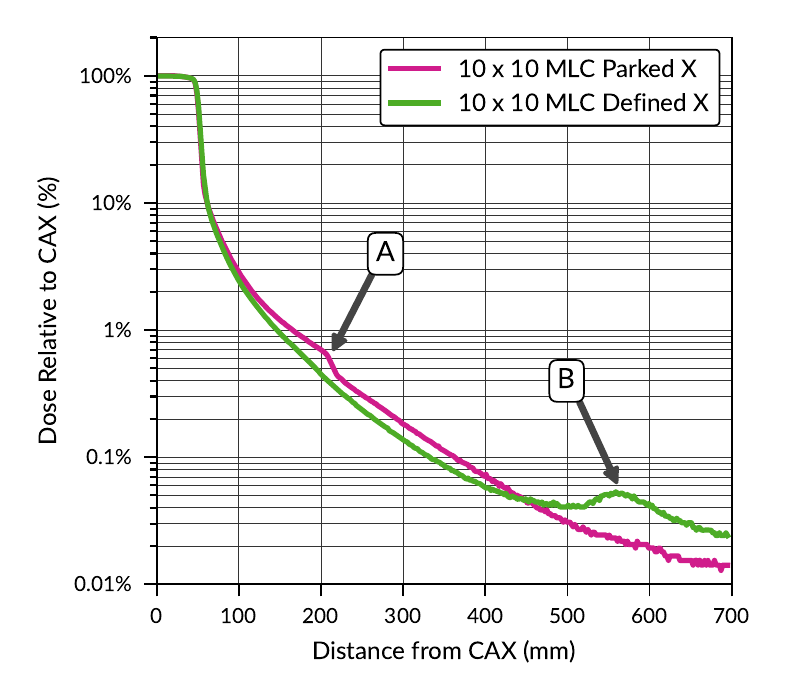}
\caption{Two 10 $\times$ 10 cm$^2$ X profiles, one MLC defined, and one jaw defined (MLC fully retracted). Marked features are explained in the article text.}\label{fig:x_profile} \end{figure}

Figure \ref{fig:x_profile} presents two 10 $\times$ 10 cm$^2$ X profiles, one of which is defined by the MLC, and the other is defined by the jaw (MLC retracted). Both profiles were measured in the transverse direction. Feature A corresponds to a sharp decrease in relative dose in the jaw defined profile due to the combined effect of the primary collimator and the MLC leaves being retracted to this position. This feature is not seen in the MLC defined profile. Feature B denotes a large difference in the profiles very far out-of-field. This may be explained by this region being shielded by the tails of the MLC leaves when they are fully retracted, but not shielded when the leaves are moved in to define the field. 

\begin{figure}[h]
\centering
\includegraphics[width=1.0\textwidth]{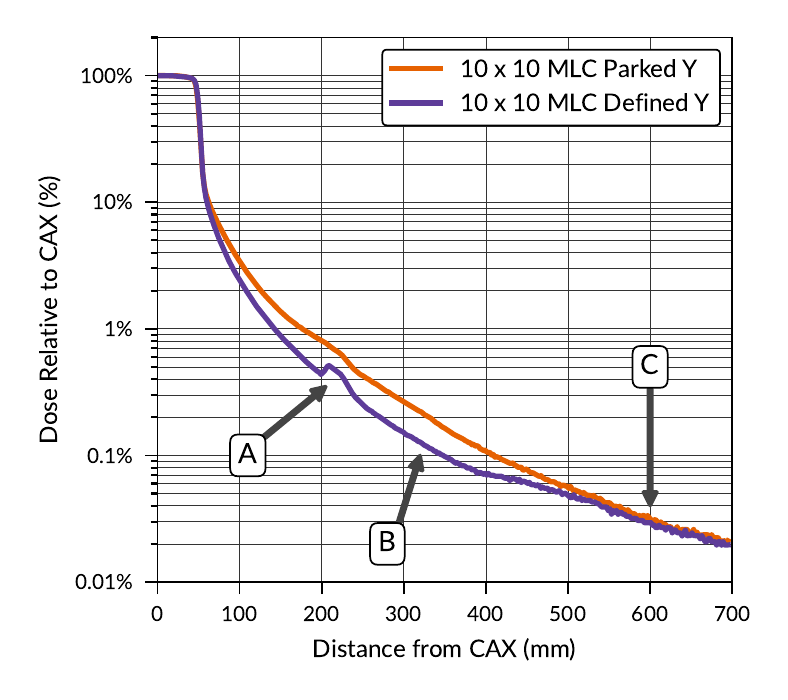}
\caption{Two 10 $\times$ 10 cm$^2$ Y profiles, one MLC defined, and one jaw defined. Marked features are explained in the article text.}\label{fig:y_profile}
\end{figure}

Two 10 $\times$ 10 cm$^2$ Y profiles, one of which is defined by the jaw (MLC retracted), and the other is defined by the MLC, are shown in Figure \ref{fig:y_profile}. Both of these profiles were collected in the radial direction. Notably, the Y jaw defined profile is the only field arrangement giving a pure jaw profile uncoupled from additional MLC shielding effects. Three features are marked in the figure. Feature A denotes a sharp increase in relative dose at approximately 20 cm off axis, due to the finite lateral extent of the MLC. This feature has been observed in an earlier study \cite{ruben2011}. This is immediately followed by a decrease in relative dose due to the primary collimator, which can also be seen on the jaw defined profile. Feature B denotes an area in which the two profiles have the same shielding conditions (jaw, primary collimator, no MLC) and yet have different relative doses. This may be explained by a difference in lateral phantom scatter from leakage more centrally in the field. The profile with both the jaw and MLC shielding the beam has less leakage relative to the jaw only profile, and therefore also has less lateral phantom scatter from this leakage. This is supported by the profiles coming into agreement in region C very far out-of-field. By this point most of the phantom scatter has been attenuated, leaving only leakage through the primary collimator and head shielding. 

\begin{figure}[h]
\centering
\includegraphics[width=1.0\textwidth]{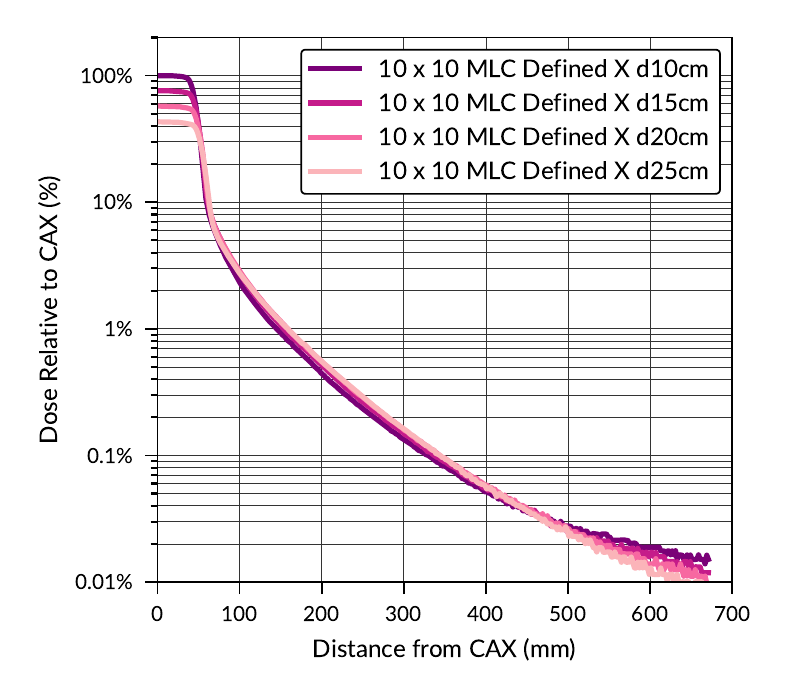}
\caption{10 $\times$ 10 cm$^2$ MLC defined X profiles measured at four different depths. All four profiles have been normalised to the central axis value of the 10 cm depth profile.}\label{fig:depth}
\end{figure}

The 10 $\times$ 10 cm$^2$ MLC defined profiles were measured at depths of 10, 15, 20, and 25 cm to investigate the variation of off-axis ratio with depth. Figure \ref{fig:depth} displays the X profiles normalised to the central axis dose of the profile at 10 cm depth. Remarkably, the out-of-field dose differs only marginally between depths. This indicates that out-of-field doses are largely independent of depth, at least at the distances and depths measured in this study. This finding is in line with earlier studies \cite{stovall1995}, but may not hold closer to the surface due to electrons scattered from the treatment head \cite{kry2006}. 

\begin{figure}[h]
\centering
\includegraphics[width=1.0\textwidth]{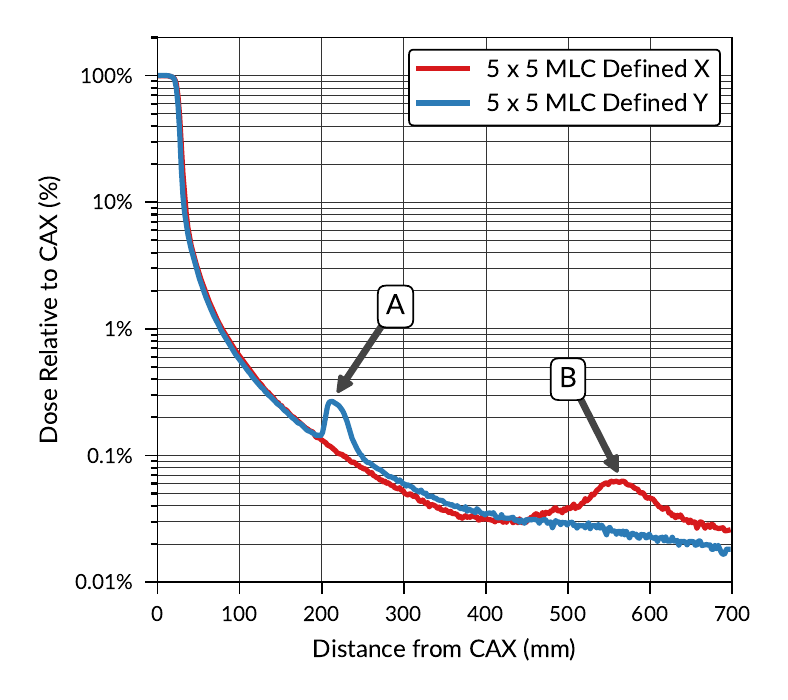}
\caption{X and Y profile of the 5 $\times$ 5 cm$^2$ MLC defined field. Marked features are explained in the article text.}\label{fig:mlc_length}
\end{figure}

As shown in Figures \ref{fig:x_profile} and \ref{fig:y_profile}, many features in the out-of-field dose are related to the MLC. Figure \ref{fig:mlc_length} shows X and Y profiles for a 5 $\times$ 5 cm$^2$ MLC defined field. The two features A and B have been seen earlier (the former related to the limited extent of the MLC bank in the Y direction exemplified in figure \ref{fig:y_profile}, and the latter related to the finite length of the MLC leaves exemplified in figure \ref{fig:x_profile}), however, the magnitude of the features relative to the central axis are larger for the smaller field compared to the 10 $\times$ 10 or 15 $\times$ 15 cm$^2$ fields. Due to the smaller field size, there is less phantom and collimator scatter to wash out these features in the relative dose.

\begin{figure}[h]
\centering
\includegraphics[width=1.0\textwidth]{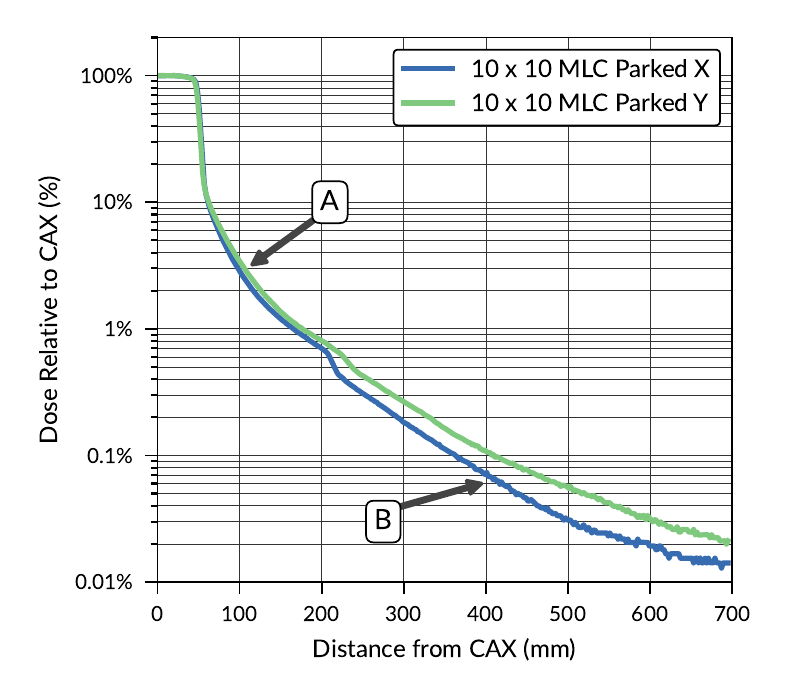}
\caption{Comparison of a 10 $\times$ 10 cm$^2$ jaw defined Y profile with an MLC defined X profile. Marked features are explained in the article text.}\label{fig:mlc_shielding}
\end{figure}

Figure \ref{fig:mlc_shielding} presents the 10 $\times$ 10 cm$^2$ X profile where the MLC is fully retracted with the 10 $\times$ 10 cm$^2$ Y profile in which there is no MLC present. From the central axis to approximately 21 cm off-axis, denoted feature A, there is a small difference in the relative dose likely due to the difference in vertical position of the X and Y jaws. Beyond 21 cm from the central axis (feature B), the presence of the retracted MLC reduces the X profile out-of-field dose to about 50\% to 75\% of the Y profile with no MLC present. This demonstrates the substantial effect that the MLC can have on shielding radiation far out-of-field, as noted by other authors \cite{mutic1999}.

\begin{figure}[h]
\centering
\includegraphics[width=1.0\textwidth]{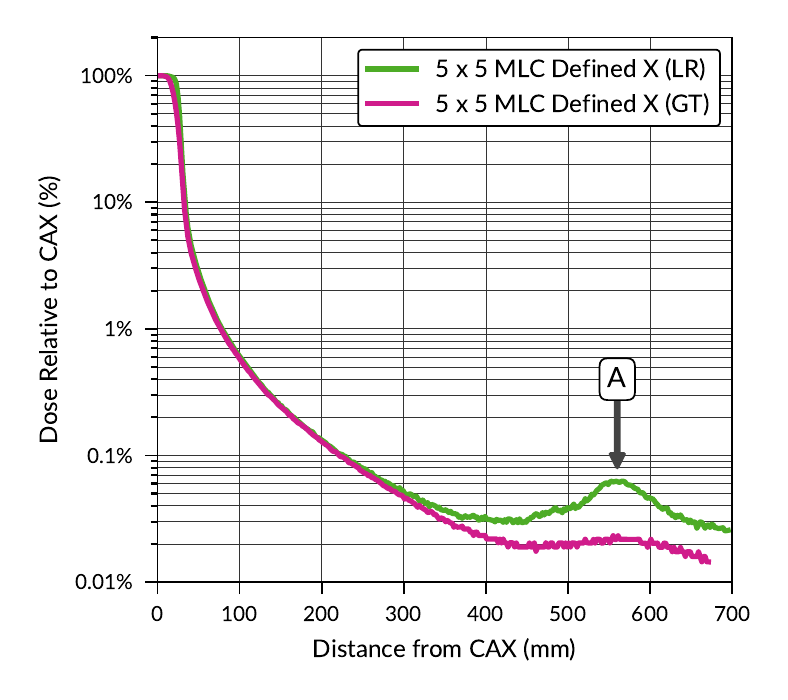}
\caption{Comparison of a 5 $\times$ 5 cm$^2$ MLC defined X profile measured in both the radial (GT) and transverse (LR) orientations. Marked features are explained in the article text.}\label{fig:trans_rad}
\end{figure}

All profiles were measured in both the radial and transverse directions. Differences were only noted very far out-of-field. In all cases, the general shape of the relative dose out-of-field was similar between radial and transverse measurements, however, the transverse relative doses were of greater magnitude. This was best exemplified in the 5 $\times$ 5 cm$^2$ MLC defined configuration, shown in Figure \ref{fig:trans_rad} (feature A). This effect may be explained by an asymmetry in the linac head shielding. 

\subsection{Limitations}

Many aspects of the above discussion involve observing features in out-of-field dose profiles and relating those features to the internal geometry of the linac head. There are many assumed relationships that cannot be stringently verified without Monte Carlo simulations that include a high-fidelity reproduction of the linac head, including the dimensions and locations of shielding blocks. Linac vendors would need to be willing to share detailed 3D models to facilitate this as the phase spaces and simplified geometries commonly shared would not be suitable for modelling complex interactions in the head shielding. Such an arrangement has been possible in the past \cite{kry2006, kry2007, bednarz2009}, and we encourage all vendors to be open to sharing this information into the future.

The measurements reported in this study were performed in a radiotherapy treatment clinic with a single model of linac, and so the extent to which this dataset can be extrapolated to other linac models is not obvious. It is reasonable to expect that the near out-of-field results may align with other models with similar tertiary collimation systems, such as the Varian Truebeam with Millennium 120 MLC. However, far out-of-field the results are more likely to depend on the exact head shielding arrangements, and so the similarity between systems in unclear. 

Linacs with markedly different collimation systems, such as those produced by Elekta (Stockholm, Sweden), may have substantially different out-of-field features. Ultimately, we recommend that clinics gather their own dataset using the techniques presented in this study. We note however that our method would need substantial modification for use with cylindrical water tanks such as the 3D SCANNER (Sun Nuclear Corporation, Melbourne, USA), but should be directly transferable to other square tanks.

\subsection{Recommendations}

We recommend that this dataset is used to further investigate and understand features of out-of-field dose distributions, particularly by assisting in the creation and validation of computational models. Furthermore, we encourage readers to perform their own measurements and compare and contrast them with this dataset. 

When planning the tank shift, we recommend that the overlap region of the inner and outer profile sections be centred around 20 cm from the central axis. This region contains several strong features, such as the beginning of the primary collimator, which allows the profiles to be stitched with confidence. 

\begin{figure}[h]
\centering
\includegraphics[width=1.0\textwidth]{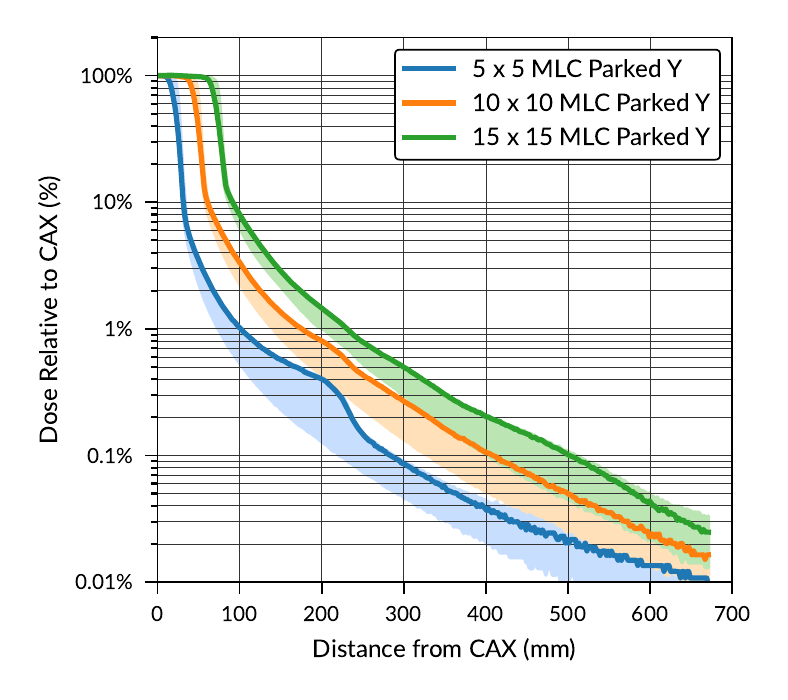}
\caption{The jaw defined Y profiles for the three field sizes in this study (solid lines). The shaded areas correspond to the complete range of profiles measured for each field size.}\label{fig:recommended_profiles}
\end{figure}

For clinics wishing to save time by collecting a more limited dataset for radiation protection calculations, we recommend collecting a single profile for a variety of field sizes. We believe the ideal profile is under the Y jaw, with the MLC fully retracted, in the radial orientation. This specific profile is compared against all others in Figure \ref{fig:recommended_profiles}. From the figure it can seen that this profile represents an upper bound of the out-of-field dose for much of the range covered in this study, and can therefore be used as a conservative estimate of the dose across this range. Being in the radial direction, it is also representative of the majority of radiotherapy treatments. 

We also make the suggestion that this same profile may be a good candidate for aiding in the development of out-of-field dose calculation models. It is a good basic test case as it is the only profile uncoupled from extra MLC shielding.  MLC shielding may then be added as a second order effect.

\section{Conclusion}

This article has presented a comprehensive dataset of dose profiles outside of the treatment field. These profiles were collected in a non-academic radiotherapy treatment centre, with standard equipment, using a technique that should be repeatable elsewhere. Ultimately, we encourage other clinics to collect their own datasets and make them freely available to the community. Access to high-quality collections of out-of-field profiles for a range of contemporary linacs would be extremely useful for risk assessments, radiation protection studies, and the development and commissioning of out-of-field dose calculation models.

\backmatter

\bmhead{Acknowledgements}

The authors would like to thank Steven R. Sylvander for his helpful insights during the planning of this study.

\section*{Declarations}

\begin{itemize}
\item Funding:\\
N/A
\item Conflict of interest:\\
All authors declare that they have no conflicts of interest.
\item Ethics approval:\\
This article does not contain any studies with human participants performed by any of the authors.
\end{itemize}


\bibliography{sn-bibliography}


\end{document}